\documentclass[11pt]{article}
\usepackage{amssymb}
\usepackage[utf8]{inputenc}
\usepackage{mathtools,tabularx}
\usepackage{orcidlink}
\usepackage{authblk}

\usepackage[backend=biber, sorting=none]{biblatex}
\hypersetup{breaklinks=true}
\DeclareUnicodeCharacter{249C}{{\scriptsize(a)}}

\renewcommand{\arraystretch}{1.3}

\begin{document}

\title{Author Intent: \\ Eliminating Ambiguity in MathML}
\author[1]{David Carlisle
\orcidlink{0009-0005-3048-4899}}
\author[2]{Paul Libbrecht
\orcidlink{0000-0003-3176-3361}} 
\author[3]{Moritz Schubotz
\orcidlink{0000-0001-7141-4997}} 
\author[4]{Neil Soiffer
\orcidlink{0000-0001-8521-1701}}

\affil{Numerical Algorithms Group Ltd, Oxford, UK \texttt{david.carlisle@nag.co.uk}}
\affil{IU International University of Applied Science, Erfurt, DE
\texttt{paul.libbrecht@iu.org}}
\affil{FIZ Karlsruhe, - Leibniz Institute for Information Infrastructure, Berlin, DE 
\texttt{moritz.schubotz@fiz-karlsruhe.de}}
\affil{Talking Cat Software Inc, Portland, Oregon, USA \texttt{soiffer@alum.mit.edu}}
%
%
%

\maketitle

\begin{abstract}

MathML has been successful in improving the accessibility of mathematical notation on the web. All major screen readers support MathML to generate speech, allow navigation of the math, and generate braille. A troublesome area remains: handling ambiguous notations such as \( \vert x\vert\). While it is possible to speak this syntactically, anecdotal evidence indicates most people prefer semantic speech such as ``absolute value of x'' or ``determinant of x'' instead of ``vertical bar x vertical bar'' when first hearing an expression. Several heuristics to infer semantics have improved speech, but ultimately, the author is the one who definitively knows how an expression is meant to be spoken. The W3C Math Working Group is in the process of allowing authors to convey their intent in MathML markup via an \texttt{intent} attribute. This paper describes that work.
\end{abstract}
\providecommand{\keywords}[1]{\textbf{\textit{Index terms---}} #1}

\setcounter{footnote}{0} 
\section{Background}

The W3C first recommended MathML as the method for including mathematical expressions in web documents in 1998. Browser adoption was slow, but by early 2023, all the major browsers supported MathML. Support for MathML by screen readers came along many years before that milestone. The quality of the speech and the number of languages supported (both in speech and in braille) varies. Some math-specific Assistive Technology (AT) software\footnote{The focus of this paper is on mathematical expressions. Accessibility of expressions involves speech, navigation, and braille. We do not discuss other important accessibility topics such as plots/graphs and diagrams.} such as SRE~\cite{cervone2019adaptable}, MathPlayer~\cite{soiffer2007mathplayer}, and MathCAT~\cite{soiffer2024mathcat} put significant effort into inferring the semantics so that superscripts are not always powers and pairs of vertical bars do not always mean absolute value. Some examples of ambiguity are given in Figure~\ref{ambiguous}.

\begin{figure}

\renewcommand\arraystretch{1.5}
\setlength\tabcolsep{7pt}
\centering
\begin{tabularx}{\textwidth}{|l|X|}
\hline
\( x^{T}\) & 
``x to the T'' or ``x transpose'' \\ 
\hline
\( x'\) & 
``x prime'' or ``first derivative of x'' \\ 
\hline
\(\overline{AB}\) & 
``line segment A B'' or ``complex conjugate of A times B''\\ 
\hline
\( (a,b)\) & 
``point a comma b'', ``open interval from a to b'' or ``gcd of a and b'' \\ 
\hline
\( x\vert y\) & 
``x divides y'', ``x such that y'' or ``x given y'' \\ 
\hline
\end{tabularx}

\caption{Examples of ambiguous notations}\label{ambiguous}
\end{figure}

One way to avoid ambiguity is to speak the expression syntactically. For example, \(\left\vert x\right\vert\) can be spoken as ``vertical bar x vertical bar'' and \(\overline{AB}\) can be spoken as ``start A B end grouping with line above.'' To our knowledge, there have not been any studies that compare listener preference for syntactic speech vs. semantic speech. It seems likely that semantic speech is preferable because syntactic speech is usually not what people are used to hearing and is also often longer. Comparing the speech for \(\left\vert x\right\vert\) and \(\overline{AB}\), syntactic speech requires 9 and 10 syllables, respectively, versus 6 and 5 syllables for semantic speech.

The most common approach to generate semantic speech is to infer what the author means by looking at the notation and its arguments. For example, to distinguish between absolute value and determinant, a single capital letter or a square table as the argument between verticals bars (e.g., \(\left\vert M\right\vert\)) would likely be a determinant, not an absolute value~\cite{schubotz2016semantification}. To varying degrees, AT looks at the arguments to generate speech for expressions like \( x^{2}\) (``x squared''). MathPlayer~\cite{soiffer2007mathplayer} goes further than most; it has over 800 patterns to improve speech. About 200 of these patterns are only active if the user specifies a subject area (this number includes chemistry speech rules). For example, if the user chooses the subject area ``probability \& statistics'', then \(\overline{x}\) is read as ``mean of x'' instead of ``x bar''. If ``calculus'' is chosen, $\times$ is read as ``cross product'' rather than ``times''. This helps resolve ambiguity at the cost of having the user inform MathPlayer about the content that is being read. MathPlayer’s use of subject area was a motivating factor for the W3C Math Working Group to explore adding ``intent''. However, it is not rare to see texts that use the same notation for different concepts, e.g. $(a,b)$ for both the open-interval and the coordinates of a point.

Another approach is to use the surrounding textual context to understand the math content~\cite{schubotz2017evaluating}. In 2017, we presented a method that extracts definitions for identifiers with an F1 score of 36\%.
Language models are rapidly improving.
In 2023, Bansal, et. al.~\cite{bansal2023extracting} described a matching learning approach to recognize definitions of symbols used in an expression by looking at the immediate surrounding context. For example, their work deduces from ``Let $F:{\mathbb R}^n\longrightarrow {\mathbb R}^n$ be a \( C^{1}\)-vector field'' that \( F \) is a \( C^{1}\) vector field. The paper from 2022 lists an F1 score of 75\% in their data set for finding a definition. However, no numbers indicate how often this is useful for improving speech, which is the final goal of their work.

For large expressions, a number of people have advocated that an overview (or outline or summary) of an expression be given. AsTeR~\cite{raman_1994} automatically elided subexpressions, but no study was done on its effectiveness.
As part of the MathGenie project, a study~\cite{gillan_barraza_karshmer_2004}  showed that providing an outline slowed solution time. Nonetheless, outlines were included in MathGenie because the authors felt it would be useful.
MathPlayer provides an option to describe an expression rather than read it. In a ClearSpeak navigation study~\cite{clearspeak_study_2017} using MathPlayer, user feedback was that outlines were not very useful. The study authors feel part of this is because the implementation was crude relative to other features.
As with reading, ambiguity can arise when summarizing expressions.

In addition to speech and braille, the ability to review a portion of an expression is important in expressions that are not simple. Most AT allows for navigation of expressions via a tree-based model, not unlike the MathML representation of the expression. In \cite{howell2023math}, the authors use a touch/tactile-based approach to navigation for people who are blind but have some small amount of residual sight (enough to resolve light/dark). This allows the users to take advantage of the physical relationships in mathematical notations (e.g., numerators are above denominators) that sighted users take advantage of. They compare their prototype for an iPhone with a tactile grid overlay to JAWS and find statistically significant benefits for touch including less frustration/effort and faster relocation of items. 
While spatial navigation is more user-friendly, the problem of ambiguous notations is still present.

In \cite{KlingenbergEtAl_DigitalLearningMathsVI_2019}, the authors note that the use of audio for reading maths textbooks is on the rise, sometimes as an alternative to braille. They point out that the right to learn to read (braille) should be supported. The paper stresses the lack of high-quality studies on the topic of mathematics learning for visually impaired.

While previous work has improved the quality of the generated speech, heuristics can never be perfect and are ultimately guesses as to what the author meant. Furthermore, even when AT knows what the author means, that doesn’t necessarily indicate how the author wants a notation pronounced. For example, 1/3 can be read as ``one divided by three'' or ``one third'' depending on what is being learned. The work presented here aims to support AT to generate better speech and, in some cases, better braille from MathML expressions.\footnote{Most braille codes are based on just the basic structure of the expression (subscripts, superscripts, etc.). They are based not on semantic meaning (index, power, etc.). There are a few exceptions to this, such as needing to know whether ``:'' is meant to convey a ratio or something else in Nemeth code. ``Intent'' can also help with braille generation in those special cases.}

\section{Author Intent in MathML 4}

The above approaches significantly improve the understandability of the generated speech. However, they are still heuristics and thus sometimes wrong. To complement these efforts, the W3C’s Math Working Group is updating the MathML standard to allow specification of how an expression should be spoken  \cite{mathml4draft11_23}. Authors can use this standard to correct heuristics, or AI researchers can evaluate the performance of their heuristics. Following the idea of correcting heuristics, the W3C’s Math Working Group decided that an approach that uses progressive enhancement is most appropriate: do not require changes; instead, allow for those notations where an author wants to make sure of unambiguity. As a rule of thumb, an author might want to enhance notations in cases when she would explain it in a classroom or at presentation, e.g., when she would write \( x^{T}\) she might say ``T means transposed''.

\subsection{Author Intent Basics}

The approach the Math WG settled on is to allow \texttt{intent} and \texttt{arg} attributes on all MathML elements. The attribute’s value has a simple, functional syntax. This syntax allows both the function head (the function name along with its properties) and its arguments to be literals, references to descendant elements, or another function. Literals can be numbers or names. See section 5.1 of \cite{mathml4draft11_23} for a full grammar. A simple example for the ``absolute-value'' concept is shown in Figure~\ref{simpleintent}.

\begin{figure}[tp]

\begin{verbatim}
              <mrow intent="absolute-value($contents)">
                <mo>|</mo>
                <mi arg="contents">x</mi>
                <mo>|</mo>
              </mrow>
\end{verbatim}

\caption{Simple intent example\label{simpleintent}}
\end{figure}

References are prefixed with \verb+$+ and some descendant of the referencing element should have a corresponding \texttt{arg} attribute value. In figure~\ref{simpleintent}, this is demonstrated with the reference ``contents''. See \ref{intent-concepts} for details.

By default, ``intent'' values should be spoken as functions are spoken, so the expression in figure~\ref{simpleintent} might be spoken as ``absolute value of x'', but AT is free to use other functional ways of speaking the ``intent'', such as a terse form (``absolute value x'') or a verbose form (``the absolute value of x''). Properties (see \ref{intent-properties}) allow for other ways of speaking an ``intent''.

In parallel with the MathML 4 recommendation, the working group is fleshing out a core list of ``intent'' concepts and ``intent'' properties with proposed speech hints.\footnote{Current working drafts are linked from~\url{https://w3c.github.io/mathml-docs/} .} This list is a reference for notations/speech that AT implementations should support. The core list is intended to cover most mathematics taught up to the university level. The list includes suggested speech in a few different languages. As a complement, an open list of ``intent'' concepts and properties is maintained; new notations are constantly created so the open list will never be complete. The open list serves both as a place where people can check to see if an ``intent'' concept has already been thought about and as a source of future additions to the core list. AT is free to implement any concepts or properties in the open lists.

\subsection{Intent Concepts}\label{intent-concepts}

The function name in an ``intent'' is referred to as the concept name in MathML~4. If the AT knows nothing about the concept name, it should be spoken as written. However, the working group's ``core'' list provides names for which AT should be aware of and for which it should provide translations for the languages it supports. Some of these concepts, such as ``fraction'' have many ways they are spoken depending on the arguments (e.g., ``one third'', ``one over x'', ``one over x all over two over x'', ``meters per second''). For someone who is blind, some of these speech patterns may include start/end words or sounds to make it clear where the fraction starts and ends; for others, these extra words or sounds may hinder comprehension. Authors rarely know their readers’ needs, so MathML 4 delegates the exact speech for core concepts to AT. If an author wants to force specific words to be used, a concept name can start with an underscore; no core names start with an underscore.

Concept names are not always a name; they can also be a reference. A reference (either the concept name or an argument) can be any child with an \texttt{arg} attribute. References start with a \verb+$+ character. The reference does not need to be unique in the document. This allows generating software to reuse templates. The algorithm for finding a reference is to do a depth-first search of the children stopping when a matching \texttt{arg} attribute value is found. If the \texttt{arg} attribute value matches the reference, the search is done. Otherwise, the element is treated as a leaf and the search continues in the parent. Figure~\ref{nestedintent} shows an example with nested ``intents'' for a nested power \(\left(x^{2}+y^{2}\right)^{2}\) that might come from software that uses a template for powers. If an \texttt{intent} has illegal syntax or references nonexistent \texttt{arg} attributes, the \texttt{intent} should be ignored by AT.

\begin{figure}[tp]
\begin{verbatim}
               <msup intent="power($base,$n)">                
                 <mrow arg="base">
                   <mo>(</mo>
                   <msup intent="power($base,$n)">
                     <mi arg="base">x</mi><mn arg="n">2</mn>
                   </msup>
                   <mo>+</mo>
                   <msup intent="power($base,$n)">
                     <mi arg="base">y</mi><mn arg="n">2</mn>
                   </msup>
                   <mo>)</mo>
                </mrow>
                <mn arg="n">2</mn>
               </msup>
\end{verbatim}
\caption{Example of nested arguments in ``intent''\label{nestedintent}}
\end{figure}

\subsection{Intent Properties}\label{intent-properties}

By default, concept names are spoken in a functional manner, but this is not always appropriate. For example,
\(\left. x^{2} \right|_{3}\) might have the concept name ``evaluated-at'' and is typically spoken as ``x squared evaluated at 3'' not as ``evaluated-at of x squared and 3''. To solve this problem, ``intent'' can be given a ``fixity'' property. The allowed values are ``function'', ``silent'', ``prefix'', ``infix'', and ``postfix''.\footnote{The Math WG is still considering adding other values such as ``matchfix'' to allow for other speech patterns.} Properties begin with ``:'' and there is no limit to the number of properties that can be attached to a concept name. For ``evaluated-at'', we might have

\verb+<mrow intent="evaluated-at:infix($expr, $value)">+ $\ldots$ \verb+</mrow>+

Early on, the Math WG realized that some notations that make use of the \verb+mtable+ element are complicated to specify using just concept names. For example, each equation in a system of equations is often divided up into columns to force alignment. To bring each equation back together, it would be necessary to list all the entries in each row as part of an ``equation'' concept. To remedy this, table properties tell AT how to speak the children. For a system of equations, the table can be marked with the ``intent'' value ``system-of-equations'' and AT should ignore the columns and just speak the table as (for example) ``2 equations, equation 1 \ldots, equation 2, \ldots end equations''. The current list of core table properties is ``matrix'', ``piecewise'', ``system-of-equations'', ``lines'', and ``continued-row''. 

Other properties are used to avoid having generating software know lots of related names. These include properties for chemical elements, units, and roman numerals.
There is also a ``chemical-equation'' property that notifies AT that chemical notation is being used so subscript, superscripts and some operators are spoken appropriately (e.g., ``='' is a double bond, not ``equals'').

\subsection{Intent and Content MathML}
In MathML, two families of elements are defined: Presentation MathML encodes how expressions are set out typographically with such typical features as subscripts/superscripts, fractions, or bracket-pairs.
Content MathML encodes how expressions are understood or interpreted with typical features as function applications, quantifiers, or externally documented symbols.

MathML expressions, i.e., semantics elements, thus can have two trees: a content tree and/or a presentation tree. Linking between content and presentation elements can be done with references or nested semantics elements. Content MathML is made available by and for computing engines. Some translations between presentation and content (with many assumptions) exist with limited scope (e.g., for simple equation expressions). While there may be an interest in finding on the web expressions with content MathML (e.g., to allow readers to perform computations), they are much less frequent than presentation MathML.

A few experiments led by members of the W3C Math Working Group have shown that content MathML can be used to generate a better accessible presentation of mathematical formulas if the semantic is available. However, the fact that content MathML is less widespread and sometimes less able to encode all mathematical discourse (without adding many symbols) has pushed the W3C Math group to propose a structure that is closer to the speech and that applies to expressions which are directly made in presentation MathML.

\section{Evaluating the Success of Intent}

While ``intent'' is aimed at improving speech for end users, the main target for ``intent'' are authors of documents that contain MathML. The reason tests for end users are not a focus is because ``intent'' should only improve the accessibility of math and never make it worse: ``intent'' gives AT the ability to \textit{know} the author’s intention of how the notation should be spoken as opposed to having to \textit{guess} the best way to speak the notation. AT has the choice to make use of this information or ignore it (e.g., to produce a syntactic reading rather than a semantic one).
Therefore, the way to measure the success of ``intent'' is two-fold:

\begin{enumerate}
	\item Are software developers that generate or consume MathML generating or consuming ``intent'' or planning to generate or consume ``intent''?

	\item For the authoring software that generates ``intent'', do users make use of facilities provided so the software can generate ``intent''?

\end{enumerate}
It is very early to evaluate either of these criteria given that the MathML 4 recommendation has not even moved to a Candidate Recommendation, let alone an actual recommendation. For the first question, there are definite signs of success as both MathML generators and AT that consumes ``intent'' have prototypes.

On the generating side, all of the prototype authoring tools that have been developed are text-based and have some resemblance to \TeX. In general, the authoring tools use macros or special syntax such as ``\verb|\abs|'' and `\verb|\det|'' to pass along the author’s intent when there is ambiguity. Discussions have also included the use of optional macro arguments to pass along intent information. For example: \verb+\times[intent=cross-product]+ would produce the following MathML: \verb+<mo intent="cross-product">+$\times$\verb+</mo>+

Three prototypes have been developed so far:

\begin{description}
	\item[\textbf{WikiTexVC}:] \cite{stegmuller2024wikitexvc} has the option to add valid ``intent'' syntax with a pseudo \TeX\ macro 
 to add ``intent'' to the MathML produced by Wikipedia, and other projects using MediaWiki. 

	\item[\textbf{UnicodeMath}:] \cite{sargent2024welcome} adds keywords that are Unicode characters when needed to resolve ambiguity and uses those to generate intents. For example, ``⒜x'' uses the Unicode code point ``⒜'' (U+249C) to indicate that what follows is the absolute value of x. This is used in speech generation.

	\item[\textbf{SpaceMath}:]\cite{farmer2024space} a superset of \LaTeX\ (both text and math) and AsciiMath that includes keywords and macros to generate intents. It will likely become an option for PreTeXt authors.

 \end{description}

No WYSIWYG editors that make use of ``intent'' have been developed yet, but ideas on how they might do this involve the use of specialized templates such as one for binomial that would generate an appropriate ``intent'' even if using the MathML-code as that of a 2d-vector. Another option is to allow users to select a symbol or expression and provide a menu of options for that symbol or expression. For example, selecting ``$\times$'' might pop up a menu with the options: ``times'', ``cartesian-product'', ``cross-product'', ``direct-product'', and ``custom\ldots''.

On the consuming side, both UnicodeMath and MathCAT implement ``intent''. Support for ``intent'' is included in the release version of MathCAT that is used in NVDA and JAWS and several other ATs. To generate speech, MathCAT first produces an intent tree from MathML. This is trivial if ``intent'' is given. If there is no \texttt{intent} attribute, then MathCAT tries to infer the intent using heuristics. The ``intent'' tree is then used to generate speech in various languages and in different styles of speech. The Math Working Group created a document with many examples comparing MathCAT’s speech with and without ``intents''. Among the lessons learned from this exercise was that ``intent'' properties for tables such as those used for aligned systems of equations greatly simplified MathML generation without complicating the implementation.

At this point in time, we lack information as to users' willingness to use features that allow generation of ``intent''. This is because the number of users of these prototypes is small and mostly includes the software authors and their colleagues. However, all have reported that generating ``intent'' is relatively straightforward. One of the PreTeXt implementers reported that authors are generally willing to improve the accessibility of their books if it is not much of a burden [D. Farmer, personal communication, 4 April, 2024]. As with many accessibility features (e.g., using headings in documents rather than just changing the font), getting users to use styles/keywords/macros likely requires education as to their benefits. All the prototypes try to minimize any extra work an author needs to do to improve the accessibility of the math.

\section{Conclusions}

The development of author intents has taught us how flexible mathematical notation can be and how this flexibility is important to mathematicians in their day-to-day practice; all mathematicians we have talked to indicate that inventing a new notation, and being able to exploit it, is common. This flexibility is supported by author intents which make it possible to encode new or existing concepts on any MathML expression  which results in speech that is accessible to the reader.

Different communities use different notations. Where these notations overlap, there is ambiguity. It does not seem possible to cleanly partition these communities because they overlap. Because of this, the only way to resolve ambiguity is at the notation level where the author tells the AT what should be said.
As an example, the last row of Figure~\ref{ambiguous} \( x\vert y\) demonstrates a notation that could be used in a number theory course with the sense of integer division, as such-that in a proof, and of conditional probability all in one paragraph.

It is important to note that ``intent''  is forward-looking: only new documents can use it. Earlier work using heuristics and textual context are still important to handle legacy documents.

\printbibliography

@article{bansal2023extracting,
title = {Extracting Contextual Semantic from a Concordance Containing Mathematical Definition},
doi = {10.3233/SHTI230607},
journal = {Studies in Health Technology},
year = {2023},
author= {Akashdeep Bansal and Volker Sorge and M Balakrishnan}
}

@article{cervone2019adaptable,
title = {Adaptable accessibility features for mathematics on the web},
journal = {16th Int. Web for All Conference},
year = {2019},
pages = {1-4},
author= {Davide Cervone and Volker Sorge}
}

@article{soiffer2007mathplayer,
title = {MathPlayer v2. 1: web-based math accessibility},
journal = {7th ACM SIGACCESS Conference},
year = {2007},
month = {10},
pages = {257-258},
author= {Neil Soiffer}
}

@manual{soiffer2024mathcat,
title = {MathCAT: Math Capable Assistive Technology},
year = {2024},
author= {Neil Soiffer},
note={Available at \url{https://nsoiffer.github.io/MathCAT/} 
\href{https://archive.softwareheritage.org/swh:1:snp:0a30a1cf4b8d62ffcdbbda33632b8dc6baca22bd;origin=https://github.com/NSoiffer/MathCAT}{swh:1:snp:0a30a1cf4b8d 62ffcdbbda33632b8dc6baca22bd}}
}

@misc{stegmuller2024wikitexvc,
title = {WikiTexVC: MediaWiki's native LaTeX to MathML converter for Wikipedia},
doi = {10.48550/arXiv.2401.16786},
year = {2024},
author= {Stegmuller, J and Schubotz, M}
}

@manual{sargent2024welcome,
title = {Welcome to UnicodeMath},
year = {2024},
author= {Murray Sargent},
url = {https://unicodemath.org/},
urldate = {2024-02-02}
}

@manual{farmer2024space,
title = {Space Math -- A Mathematical Notation System \& \LaTeX\ Translator},
year = {2024},
author= {David Farmer},
url = {https://github.com/davidfarmer/SpaceMath},
urldate = {2024-02-02},
note = {\href{https://archive.softwareheritage.org/swh:1:snp:fdc8ee5b7404ae927d4bd6fa14c3d0c6d0cf9781;origin=https://github.com/davidfarmer/SpaceMath}{swh:1:snp:fdc8ee5b7404ae927d4bd6fa14c3d0c6d0cf 9781}}
}

@article{howell2023math,
title = {Math for Those with Severe Low Vision: From the Particulars to the Gestalt (And Back Again)},
doi = {10.1109/FIE58773.2023.10343509},
journal = {IEEE Frontiers in Education Conf.},
year = {2023},
pages = {1-9},
author= {Howell, J and Quek, F}
}

@article{schubotz2016semantification,
title = {Semantification of Identifiers in Mathematics for Better Math Information Retrieval},
doi = {10.1145/2911451.2911503},
journal = {SIGIR '16},
publisher = {ACM},
year = {2016},
pages = {135-144},
author= {Schubotz, M and Grigorev, A and Leich, M and Cohl, H and Meuschke, N and Gipp, B and Abdou, S and Youssef, V and Markl}
}

@article{schubotz2017evaluating,
title = {Evaluating and Improving the Extraction of Mathematical Identifier Definitions},
doi = {10.1007/978-3-319-65813-1_7},
journal = {LNCS},
volume = {10456},
year = {2017},
publisher = {Springer},
author= {Moritz Schubotz and L Krämer and N Meuschkeh and F Hamborg and B Gipp  and G Jones}
}

@misc{mathml4draft11_23,
title = {Mathematical Markup Language (MathML) Version 4.0, Editors Draft},
year = {2023},
month = {11},
author= {David Carlisle}
}

@article{KlingenbergEtAl_DigitalLearningMathsVI_2019,
  author={Klingenberg, Oleg and Holkesik, A and Augestadt, L},
  title={Digital learning in mathematics for students with severe visual impairment: A systematic review},
   year={2019},
    volume={38},
    issue={1},
    doi={10.1177/026461961987697},
    journal={J. of Visual Impairment},
    publisher={Sage Journals}
}

@InProceedings{gillan_barraza_karshmer_2004,
doi={10.1007/978-3-540-27817-7_94},
author="Gillan, Douglas J.
and Barraza, Paula
and Karshmer, Arthur I.
and Pazuchanics, Skye",
title="Cognitive Analysis of Equation Reading: Application to the Development of the Math Genie",
booktitle="Computers Helping People with Special Needs",
year="2004",
pages="630--637",
}

@phdthesis{raman_1994,
    author = "Raman, T. V.",
    title ="Audio system for technical readings",
    school = "Cornell University",
    year = "1994"
}

@techreport{clearspeak_study_2017,
    author = "Frankel, Lois and Brownstein, Beth and Soiffer, Neil",
    title = "Expanding audio access to mathematics expressions by students with visual impairments via MathML",
    institution = "ETS",
    year = "2017",
    doi = {10.1002/ets2.12132}
}

\end{document}